# Position reconstruction in fission fragment detection using the low pressure MWPC technique for the JLab experiment E02-017

QIU Xi-Yu[1;1)]   L.Tang[2,3]   A.Margaryan[4]   HU Bi-Tao[1]   CHEN Xi-Meng[1;2)]

[1] School of Nuclear Science and Technology, Lanzhou University, Lanzhou, 730000, China
[2] Department of Physics, Hampton University, Hampton, VA, 23668, USA
[3] Thomas Jefferson National Accelerator Facility, Newport News, VA, 23606, USA
[4] Yerevan Physics Institute, Yerevan 375036, Armenia

**Abstract**

When a Λ hyperon was embedded in a nucleus, it can form a hypernucleus. The lifetime and its mass dependence of stable hypernuclei provide information about the weak decay of Λ hyperon inside nuclear medium. This work will introduce the Jefferson Lab experiment (E02-017) which aims to study the lifetime of the heavy hypernuclei using a specially developed fission fragment detection technique, a multi-wire proportional chamber operated under low gas pressure (LPMWPC). Presented here are the method and performance of the reconstruction of fission position on the target foil, the separation of target materials at different regions and the comparison and verification with the Mote Carlo simulation.

**Introduction**

Since its' discovery in 1952 [1], the Λ hypernuclei, in which one or more nucleon of the nucleus was replaced by Λ hyperon, has been studied very extensively. One of the study interests is the weak decay of the Λ hypernuclei. Generally most of the produced Λ hypernuclei stand in their excited states and then can reach the ground state by emitting nucleon or through electromagnetic decay [2]. At ground state Λ hypernuclei decay via $\Delta S = 1$ weak interactions.

In free space, a Λ hyperon decays only via mesonic weak decay into a nucleon and a pion:

$\Lambda \rightarrow p + \pi^- + 37.8$ MeV (63.9%)

and

$\Lambda \rightarrow n + \pi^o + 41.1$ MeV  (35.8%).

When a Λ hyperon is embedded into a nucleus, which forms a hypernucleus, this mesonic decay channels are significantly suppressed as the hypernuclear mass increases. This is because that the 40 MeV released energy (mass deference between the Λ mass and the summed mass of the pion and nucleon) is too small in making the decay nucleon (proton or neutron) escaping from the nucleus from the s orbit. Therefore, the non-mesonic decay mode becomes a dominant decay channel for the hypernuclear decay. This decay mode has three sub-channels:

Proton induced decay:

$\Lambda + p \rightarrow n + p + 176$ MeV (with branching ratio, $\Gamma_p$)

Neutron induced:

$\Lambda + n \rightarrow n + n + 176$ MeV ( with branching ratio, $\Gamma_n$)

Three body decays:

$\Lambda + N + N \rightarrow n + N + N + 176$ MeV (with branching ratios, $\Gamma_{NN}$), where N is either a p or n.

The 176 MeV released energy is simply the mass difference between Λ and decay nucleon and is sufficiently large for the nucleon to escape from the nucleus. Therefore, in contrast to mesonic decay, the non-mesonic decay quickly becomes dominant as mass increases. For heavy hypernuclei, the none-mesonic decay can also trigger nuclear fission or fragmentation. Therefore, time delayed fission contains characteristics of the non-mesonic weak decay of heavy hypernuclei. Based on this idea, the COSY-13 collaboration studied the lifetime of heavy hypernuclei produced by proton interaction with Au, Bi and U nuclei using a recoil shadow method [3-5]. In this method the lifetime was converted from the position distribution in the shadow region where only fragments with time delayed decay can reach. However, the conversion depends on the modeling of the momentum and mass distributions of decay hypernuclei. The resulted lifetime was quite short, ~145ps, and could not be explained by current theories on baryonic weak interactions.

The Jefferson Lab (JLab) experiment E02-017, using photo-production to produce Λ hypernuclei, aims to directly measure the lifetime of the heavy hypernuclei by utilizing the same time delayed fission characteristics as COSY-13 did but applying the low pressure MWPC technique developed in detecting both the time and position of fission fragments. Multiple target materials were spotted as strips on a single thin Mylar foil in order to simultaneously produce hypernuclei with different masses. The production rate from different targets depends on the Gaussian type of spread of the photon beam intensity distribution. This work will give a brief introduction of the experimental setup and present the method of the fission position reconstruction and the target separation result.

**Experiment**

The JLab experiment E02-017 was carried out in the experimental Hall C parasitically with the hypernuclear mass spectroscopy experiment E05-115. The entire experimental setup is illustrated in Fig.1.

The JLab accelerator can provide high quality continuous wave (CW) electron beam, which has 1.67ps pulse width and 2ns pulse separation. The beam energy used by E05-115 was 2.5GeV and beam hit the E05-115 target that located at the entrance of the splitter magnet that bent the oppositely charged scattered electrons and $K^+$'s to the following spectrometers, HES (e') and HKS ($K^+$), respectively. The post electron beam was also bent and aimed directly to the electron beam dump. The real bremsstrahlung photons produced from the E05-115 target were not affected by the splitter magnet and went straight forward straightly to a local photon dump about 15 meters downstream of the E05-115 target along the incoming beam direction. This dumped photon beam was utilized as real photon beam with a wide range of beam energy up to 2.5 GeV for the E02-017 experiment that was setup just in front of the local photon dump. The threshold photon energy to produce Λ hyperon is about 700MeV.

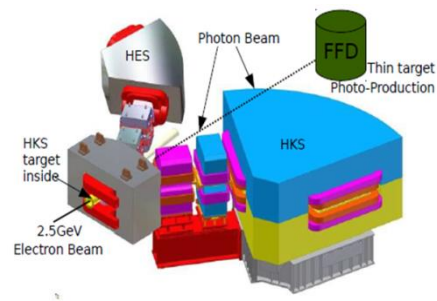

Figure 1. The experimental setup for E05-115 and E02-017 in JLab Hall C

The entire experimental system for E02-017 was contained in one common low pressure chamber filled with pure heptane gas with the pressure regulated at about 3 Torr. A thin tilted target foil was located at the center of the chamber and aligned to intercept the incoming photon beam. The fission fragment detector (FFD) system used for E02-017 was symmetrically allocated above and below the tilted target foil.

The specially developed FFD system for this experiment is based on the low pressure multi-wire proportional chamber (LPMWPC) technique [6, 7]. The detector has a modular structure and contains four independent LPMWPC units. Two units are above the target foil forming the top telescope and time-of-flight (TOF) pair and two units are below the target foil to form the bottom pair. The top and bottom pairs are completely symmetric and all detector planes are parallel to the beam direction.

Each LPMWPC unit consists of one central anode wire plane (A) at a potential of +300V, two cathode wire planes (C) at ground potential that sandwiches the anode plane and two guard planes (G) at a potential of -100V further outside the cathode planes. Thus each unit has symmetric G-C-A-C-G plane order. The A plane is formed by 12μm gold-plated tungsten wires with 1mm spacing. The wires are electrically connected together to provide a single timing signal when detecting one passing fragment. The C planes are formed by 40μm gold-plated tungsten wires also with 1mm spacing. However, wires are electrically grouped by three wires and each group connected to one tab of the digital delay line chip with 2ns delay between two adjacent tabs. A chain of digital delay line chips are mounted on one side of the board. Induced charges at a specific wire group give two signals traveling through the delay line chains in two directions, to the left (L) and to the right (R). The timing was analyzed in reference to the same A plane timing. The time difference, L-R, provides position information while the time sum, L+R, can be used to filter the noise and multiple fragment detection and be used to estimate the position resolution (or uncertainty). The two cathode planes have the wires laid normal (90°) to each other so that two dimensional position can be determined for single detection point by each unit while the A plane provides the detection time. The two G planes provide second step ionization to increase the signal size and set equal potential between units to minimize noise charges. The plane separation is 3.175 mm.

The two outer units have active area of 21 x 21 $cm^2$ while that for the inner two units is 10.5 x 10.5 $cm^2$. The vertical distance between the beam line centroid to the inner A plane is 3 cm and that between the two A planes of the inner and outer units is 7 cm. Therefore, detection of fission fragment at any unit includes information of time (T) and three dimensional positions (X, Y, Z). In our FFD geometry definition, X is in beam direction, Y is normal to X and parallel to the LPMWPC planes and Z is normal to both beam and LPMWPC planes. T is measured relative to the event trigger time. All measured parameters are labeled 1-4 from bottom unit upward. Since no particle from production process was available, the time signal from the bottom outer A plane was used as time reference for all events and therefore T1 is always a constant.

Under low gas pressure, the ionized particles have long free mean path length. The gas gain is very low in comparison to conventional MWPC or drift chambers. Thus only fission fragments with high Z can produce signals with sufficient pulse size. Low mass and low Z fragments or particles cannot be detected. Since the signal is still at least 10 times smaller than that from regular MWPC, each signal is pre-amplified by 100 times with a fast pre-amplifier mounted on the edge of the associated wire plane. The analog signals are then extracted outside of the vacuum sealed chamber and amplified 10 times again before converting to logical signals. More detailed description of this FFD can be found in Ref. 6 and 7.

The target used in the experiment was made by sputtering various materials onto a 2.0 μm thick aluminized Mylar foil. The Submitted to "Chinese Physics C".

Mylar functioned as the backing support and was stretched over a rectangular aluminum ring frame with an active area of $13.2\times 8.2 \mathrm{cm}^2$. The Fe target was specially prepared in order to make consistency verification with the previous measurement done by KEK group on $_\Lambda$Fe result. The target foil surface was placed in a small incline angle (10°) with respect to the beam direction. This was to maximize the target thickness for the beam photons in X-direction, i.e. increasing the production rate, while to minimize the thickness for the escaping fission fragments towards the LPMWPCs, i.e. maximizing the fragment escaping rate.

Table 1 lists all the materials prepared for the experiment, the thickness and the actual separations between materials. The list follows the same sequence as the material arrangement on the backing foil and Fe material is at the center with respect to the beam. The small incline angle allowed the photon beam with narrow angular distribution to cover wider range of target materials. However, the sharp beam distribution still did not allow productions from Au and U targets. Thus, data were collected only from the four targets closer to the beam center. In addition, 2.5 mm separation was originally planned between target regions but the masking difficulty during the sputtering process made smaller and uneven separations. In case of Cu and Au, the mask was completely failed. More discussions will be given later in the events separation.

In addition, a 252Cf spontaneous fission source is mounted in a source holder with a collimator at a distance of 13.73cm from the anode plane of the top unit. The fission fragments escaped from the source were used to test FFD. Source test results and characteristic performance without target foil and beam can be found in Ref. 6. During the experiment, the source events were also used for checking the condition of FFD. However, the three extra foils (the target foil and the two beam induced charge separation foils made by the same 2 μm aluminum Mylar) made significant difference between the velocities measured by the top pair and bottom pair of LPMWPCs due to energy loss which is a function of momentum. Therefore, under experimental condition with foils inserted, the fragments from source were used only to check basic operational conditions and alignment calibration. The structure and geometry of FFD discussed above is schematically illustrated in Fig. 3.

Table 1. Target materials used in the experiment.

| Target material | Thichness/μm | Width/mm | Separation/mm |
|---|---|---|---|
| Au | 0.4 | 29.0 | |
| | | | 0.0 |
| Cu | 0.8 | 25.5 | |
| | | | 2.0 |
| Fe | 0.8 | 20.0 | |
| | | | 1.0 |
| Ag | 0.8 | 21.5 | |
| | | | 1.0 |
| Bi | 0.4 | 20.0 | |
| | | | 1.5 |
| $U_{natural}$ | <0.003 | 11.5 | |

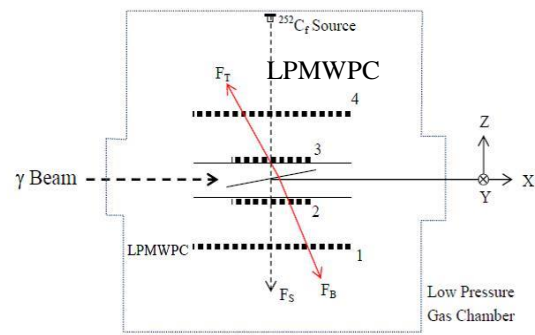

Figure 3. Schematic illustration of the structure and geometry of FFD.

**Position reconstruction method**

The position where a fragment passed through a LPMWPC unit is determined by using the induced signals detected from the C planes, as mentioned in the previous section. The delay line technique [8, 9] is applied so that each charge induction at a detection LPMWPC unit splits into two pulse signals in

4
Submitted to "Chinese Physics C".

opposite directions. This pair of signals is converted into two times (for instance, $X_iL$ and $X_iR$ while i = 1-4 is the LPMWPC unit number) by time-digital-conversion (TDC) in reference to the anode (A) plane signal time. Then the detected position is determined by, for example, $X_i = F \times (X_iR - X_iL)$ in X direction on the $i_{th}$ unit with R is in the positive beam direction. Here F is the time to length conversion factor which is simply the total delay line time divided by the total C plane width. It needs to be calibrated when taking into account the TDC conversion factor and the precision of delay line chips. The sum of $X_iR$ and $X_iL$ (i.e. $SX_i = X_iR + X_iL$) is also an important parameter. First of all, $SX_i$ for single fragment detection should be a constant. Therefore, a tight gate in the analysis to select the events in the sharp peak appeared in the $SX_i$ distribution can be a powerful cut to eliminate background and mixing of multiple fragments. Secondly, $SX_i$ can be used to evaluate the position resolution since the uncertainty is at the same level for both ($X_iR + X_iL$) and ($X_iR - X_iL$).

Position reconstruction and the geometry alignment were verified by checking the tracking of the fragments emitted from the calibration $^{252}$Cf source mounted at the top as illustrated in Fig. 3. With the same analysis cuts above, events with single fragment penetrated through all four units were selected. The peak width of the single plane residuals, which were obtained by fitting a straight line from the four measured X positions, was about 1mm for all four units see Fig. 4. This is expected to be worse than the first test result in Ref. 6. since two charge separation foils and target foil were then inserted for the experiment. The fragments from the $^{252}$Cf source must penetrate all three foils in order to reach the bottom pair of LPMWPC. The significant energy loss and multiple scatterings increase the residuals from a straight line fit. Since the geometric information from the two dimensional X-Z planes is sufficient to make TOF reconstruction, no analysis in Y direction.

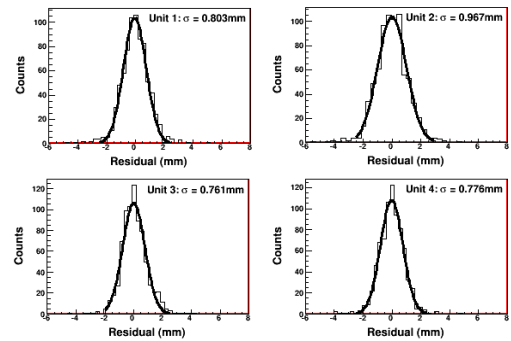

Figure 4. The single plane residuals of the four X planes

With beam on target, it is anticipated that large amount of low energy electrons induced by the beam may be piled in the target region. Its density decreases as inversed square distance. This charge background does not generate signals to be detected as fragments but reduce the electric potential between the C and A planes of LPMWPC. Therefore, the two inner units which have their A planes only 3cm away from the beam center line are expected to have less efficiency and worse single plane position resolution, i.e. a large time jitter in the signals for $X_iR$ and $X_iL$ which were used to find the position on $i^{th}$ plane. This was checked by the sum of left and right delay time, i.e. $X_iR + X_iL$ as discussed previously. Fig. 5 shows this sum for each of the X planes.

The two outer X planes have the position resolution of ~0.35-0.37 mm. This agrees with the conclusion from the source study in Ref. 6. Since they are 10cm away from the beam, the piled charge effect was almost negligible. However, the inner units were affected by this charge pile up obviously and significantly. The position resolution deteriorated to about 1.44 mm and 1.74 mm for unit #2 (just below the beam) and unit #3 (just above the beam), respectively. The poor position resolution of the inner units made the separation of target materials could not be done cleanly. Therefore, Monte Carlo simulation is necessary in helping to obtain the possible mixing percentages for the events from the adjacent target regions.

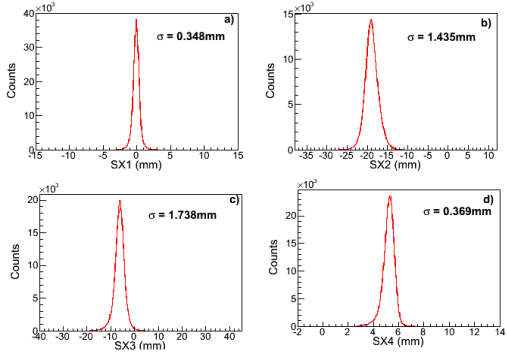

Figure 5. Sum of "Left" and "Right" delay times for each X planes in the associated LPWMC units. SX1 and SX4 are the planes in the outer two units while SX2 and SX3 in the two inner units which are close to the beam (see Fig. 3).

**Monte Carlo simulation**

In the Monte Carlo simulation, binary fission at its center of mass (CM) was assumed without lifetime. The statistical distribution of the total kinetic energy was based on the measured quasi-free mass distribution in the kinematics acceptance corrected spectrum of the $^{28}_{\Lambda}$Al hypernuclei obtained by the electro-production (JLab experiment E01-011) at small forward electron scattering angles. The change of distribution for heavier hypernuclei was assumed not affecting the simulation for position reconstruction study.

The simulated two fission fragments were generated from the target foil based on the target material regions with the measured gaps between regions. An emission angle with respect to the beam direction (+X) was generated for each of the two fragments from one fission event, using a three dimensional Monte Carlo with the probability obtained from the real measurement. Fig. 6(a) shows the correlation between the emission angles of the two simulated fragments while Fig. 6(b) is that from the real experimental measurement which was also used to extract the probability distribution function. The actual active area size of the LPMWPC units had effect to the boundary shape. A general agreement was sufficient for the purpose of the simulation study.

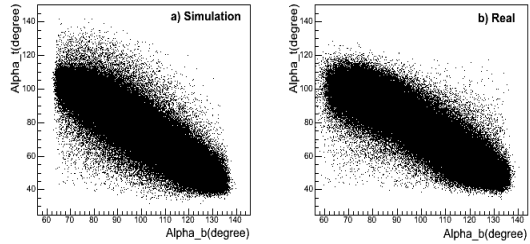

Figure 6. Emission angle correlation between the two detected fragments, (a) is generated by Monte Carlo while (b) was from the real data. The density of the distribution represents the probability.

The sum of the two emission angles is the opening angle between the two fragments. The distribution in red color in Fig. 7 shows the distribution of the measured opening angles. The blue colored distribution was that from the simulated events. Good agreement can be seen by comparing these two distributions. Single plane position resolution had small effects to mainly the shaping in the tails of the opening angle distribution.

The simulation simply used the extracted probability functions from the real measurement without identification of their physics origins. However, one feature from the opening angle distribution can be easily seen. If CM of fission is at rest, the two fragments from a pure binary fission should be emitted back to back, i.e. with an opening angle of 180 degrees. Position resolution of the LPMWPC units can broaden the opening angle distribution for only few degrees according the geometry. Due to momentum transfer in the photo-reaction, CM of fission has momentum which boosts the two fragments slightly forward in the direction of beam. Including the range (250-500 MeV/c) of 3-momentum transfer in photoproduction of heavy hypernuclei, the opening angle of binary fissions is found centered at ~177°, agreed closely with that from the measurement but the range is still small. It is commonly known that 2 to 10 nucleons can be emitted during fission of heavy nuclei. However, the tail of the opening angle distribution extended about 40° beyond the simulation that included the nucleon emissions. This indicates two additional possible fission processes. The first is from the one-step fissions that had multiple (>2) fragments with only two detected within the acceptance. The second is originated from events experienced two-step fissions. In case of heavy hypernuclei, the initially excited hypernuclei can have masses above nuclear breakup thresholds so that they decay first via break up reaction that emits a fragment to be stabilized into lighter hypernuclei (i.e. hyperfragments). The lighter hypernuclei then decay via non-mesonic weak decay by emitting two nucleons and a nucleus in the process. Thus the two fragments in this case are not from the same kinematic reaction. This reaction is the source in this experiment to study lifetime of hyperfragment and this part will not be discussed in this paper.

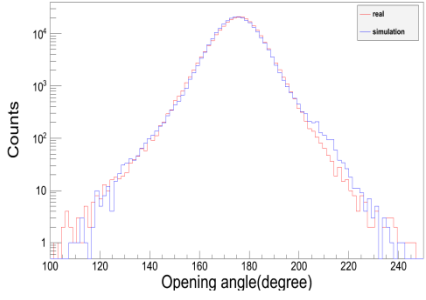

Figure 7. Distribution of the opening angles between two detected fragments. Histogram in red is from real data while the simulation is in blue.

The two Monte Carlo generated fragments had their positions calculated at the LPMWPC units according to the geometry. To study the effect from the position uncertainty, the single plane position resolution was assumed to be that from the corresponding $SX_i = X_iR + X_iL$ (see Fig.5). However, it was noticed that the peaks are not perfectly symmetric, especially in the tail regions, and could not be perfectly fitted by a single Gaussian function including the full tails. In fact, the measured $X_iR$ and $X_iL$ were originated from the same induced charges but traveled through two parts of the total delay line. Statistically $SX_i$ shows the same average uncertainty as that of $X_i = X_iR - X_iL$. But $SX_i$ and $X_i$ can have different asymmetry and tail distribution. Therefore, additional information was needed to obtain more realistic single plane resolution function for simulation purpose.

The position on the target in X direction was reconstructed separately by the two fragments detected by the top and bottom pairs. The difference (ΔX) between the two reconstructed X values ($X_{top}$ and $X_{bottom}$) is the only available additional information that can be used to study the single plane resolution, as shown in Fig. 8. Its shape is defined by the FFD geometry and reconstruction to a tilted target foil but its asymmetry contains contributions from the asymmetry tail shape of the single plane resolution function of the four LPMWPC units. On the other hand, a large ΔX does not means a large error in determining the fission position, since the fission position is decided by the mean of the $X_{top}$ and $X_{bottom}$. Uncertainty of the reconstructed fission position is determined by the uncertainty of $X_{top}$ and $X_{bottom}$. In Fig. 8, the distribution in red color is from the real data while the blue one from the simulation. Initial position resolution function for each LPMWPC unit was extracted from $SX_i$, using three Gaussian fit. The main disagreement was asymmetry shape in the tail region. The resolution function for each unit was then adjusted based on the level of sensitivity to the shape of this distribution. The final simulation result (shown in blue color) still has small residual disagreement in the tails but contains less than 3% of events. This small disagreement might come from other unknown factors. The current level of simulation is sufficient to study the fission position reconstruction and mixing of the events from adjacent target regions after region separation.

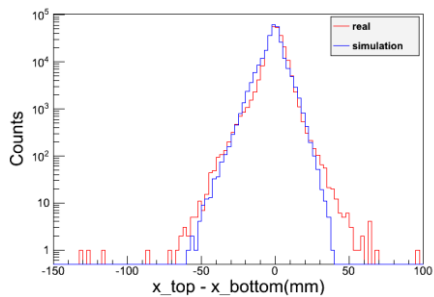

Figure 8. The difference between the target position in X direction reconstructed by the top and bottom pairs of LPMWPC units.

**Fission position reconstruction**

The fission position on the target, projected to X direction, was reconstructed by solving the intersection of the trajectory of the fragment and the target plane. This position was reconstructed separately by the fragments detected by the top and bottom pairs of LPMWPC units. The average of these two was used as the fission position on the target. In Fig. 9, the distribution with red color is the reconstructed fission position. The mean separation boundary of the target regions is marketed by the straight lines with target material labeled. Events from the Au and U materials are too rare thus they are not labeled but their information can be found in Table 1.

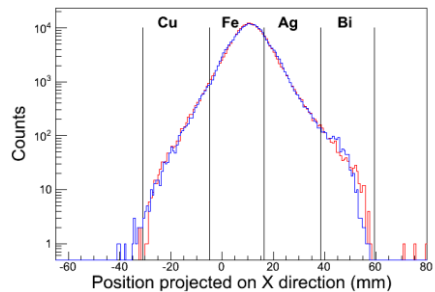

Figure 9. Reconstructed fission position on target (projected in X direction). Distribution in red is from real data while that in blue is from simulation.

It is obvious that this reconstruction distribution does not show gaps or "dips" that indicate the material separations. This makes the X coordinate alignment difficult to be verified. Also, it means that mixture of events from different target materials is unavoidable. Thus simulation is needed to verify that the absence of gaps or "dips" is resulted from limited position resolution as discussed previously and to provide an estimation of the percentage event mixing which must be taking into account when lifetime is extracted from the decay time spectrum after events separation.

The shape of this fission position distribution is dominantly determined by the photon beam intensity profile in the plane normal to the beam direction. Although fission probability increases as target mass increases, its effect is too small in competing with the sharp falling intensity of the beam. Thus, fission probability variation was not considered separately in the simulation.

Although the electron beam energy used by the hypernuclear mass spectroscopy experiment E05-115 was almost monochromatic, the photons radiated from its target had full range of energy, up to its maximum. The angular distribution of Bremstrahlungs varies according to the photon energy. Therefore, the intensity distribution is not expected being able to be described precisely by a single Gaussian function. It has significant tail distribution. In addition, the radiated photon beam was transmitted within a vacuum pipe with small diameter. Small emission angle misalignment for this experiment was observed by the asymmetry of the counting rate and efficiency between the top and bottom LPMWPC units. This is due to the tuning for the specific beam position by the experiment E05-115 for the optimized beam position on its target which is the radiator for this lifetime experiment E02-017. This misalignment caused background photons from scatterings in the beam pipe material. On the other hand, the beam spread could help to enlarge the photon beam profile size so that more target materials could be in beam simultaneously. Therefore, it was accepted. Therefore, such asymmetric distribution was expected. Unfortunately, the E05-115 target was so thin that radiation length was too short to spread the photon beam further to have events at least from the Au target. The U target material was too thin to be considered.

The distribution was fitted by using a function that contains four Gaussian functions allowing different means and widths. The extracted beam profile function gives a continued distribution with its full tails. The function is used as the probability function in the simulation in generating fission events from target foil according to its realistic geometry. No material was assumed in the U region. The distribution in blue color in Fig. 9 is the reconstructed fission position from the simulated events.

It is obvious for the sharp cut off for the number of events at the U boundary. No significant amount of events could be obtained from the Au target. Indeed, it verified that only Cu, Fe, Ag and Bi have sufficient number of events for the later lifetime study. Also, with the obtained single plane resolutions no "dips" could be seen from the simulated events.

To find a positive signature that can further verify the X alignment (besides the U boundary), $\Delta X$ gate was studied in hoping to select events with better position resolution. Fig. 10 shows the resulted fission position distribution from events within an extremely tight gate, -0.1 mm < $\Delta X$ < 0.1 mm. Only very small amount of events left. There is at least a sharp slope at the same boundary between Fe and Ag in both the real and simulated data. This helped to verify that the alignment of X is correct. On the other hand, it shows that the position resolution does not have significant contribution to $\Delta X$, as previously mentioned. It cannot be used for the purpose of improving resolution without suffering dramatic statistics loss.

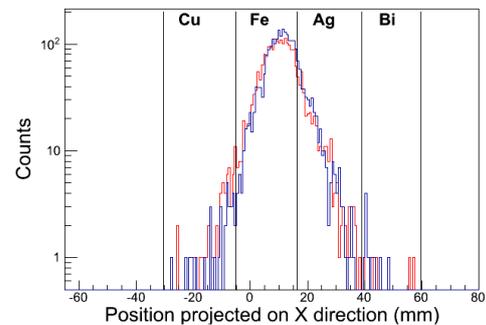

Figure 10. The reconstructed fission position on target with an extremely tight $\Delta X$ cut: -0.1mm < $\Delta X$ < 0.1mm.

**Extracted mixture of events**

Aided by the simulation, the mixture of events after target separation could be achieved. Ignoring the gap size, simple boundary was used as indicated in Fig. 9 and 10 to separate the events from different target.

When this separation was applied to the simulated events, the percentage mixture from adjacent targets is obtained, as shown in Table 2.

Table2. Events mixture in a certain target region

| Mix from \ Region | Cu | Fe | Ag | Bi |
|---|---|---|---|---|
| Cu |  | 0.53% |  |  |
| Fe | 34.6% |  | 43.9% |  |
| Ag |  | 1.6% |  | 24.13% |
| Bi |  |  | 0.16% |  |

The decay time spectrum will be analyzed for events from each separated target region. For the region that mixture is significant, multiple lifetimes with known ratio of number of events may have to be included in fitting the decay time. For the regions in which the mixture is minor, for example less than few percent, it may be treated simply as part of the overall systematic error for the lifetime.

**Summary**

In this work, the concept and the technique of the JLab experiment E02-017 which aims to measure the lifetime of the heavy hypernuclei were introduced. The crucial step in extracting the lifetime of the hypernuclei from the massive experimental data is to find the fission points on the target foil and to separate the events to the corresponding target materials. Due to limited position resolution, a substantial Monte Carlo simulation work was needed to aid the analysis. With the separated events and known mixture information, the lifetime of hypernuclei will be extracted and its final result will be reported later.

**Acknowledgement**

We thank the great support provided by the technical staffs from Accelerator Division and Hall C in Physics Division at JLab. The experimental work is partially supported by various grants that include the support received from the National Nature Science Foundation of China grants 11175075, 11135002 and 91026021 and the State Scholarship Fund program of the China Scholarship Council.